\documentclass[aps,prb,
 amsmath,amssymb,twocolumn, groupedaddress]{revtex4-1}
\usepackage{natbib}
\usepackage[dvipdfmx]{graphicx}

\usepackage{color}
\usepackage{comment}

\begin{document}

\title{Charge order with unusual star-of-David lattice in \\ monolayer NbTe$_2$}

\author{Taiki Taguchi,$^1$ Katsuaki Sugawara,$^{1,2,3,4}$ Hirofumi Oka,$^2$ Tappei Kawakami,$^1$ Yasuaki Saruta,$^1$ Takemi Kato,$^1$ Kosuke Nakayama,$^{1,4}$ Seigo Souma,$^{2,3}$ Takashi Takahashi,$^{1,2,3}$ Tomoteru Fukumura,$^{2,3,5}$ and Takafumi Sato,$^{1,2,3,6}$
}

\affiliation{$^1$Department of Physics, Graduate School of Science, Tohoku University, Sendai 980-8578, Japan}
\affiliation{$^2$Advanced Institute for Materials Research (WPI-AIMR), Tohoku University, Sendai 980-8577, Japan}
\affiliation{$^3$Center for Science and Innovation in Spintronics (CSIS), Tohoku University, Sendai 980-8577, Japan}
\affiliation{$^4$Precursory Research for Embryonic Science and Technology (PRESTO), Japan Science and Technology Agency (JST), Tokyo 102-0076, Japan}
\affiliation{$^5$Department of Chemistry, Graduate School of Science, Tohoku University, Sendai 980-8578, Japan}
\affiliation{$^6$International Center for Synchrotron Radiation Innovation Smart (SRIS), Tohoku University, Sendai 980-8577, Japan}

\date{\today}

\begin{abstract}
Interplay between fermiology and electron correlation is crucial for realizing exotic quantum phases. Transition-metal dichalcogenide (TMD) 1$T$-TaS$_2$ has sparked a tremendous attention owing to its unique Mott-insulating phase coexisting with the charge-density wave (CDW). However, how the fermiology and electron correlation are associated with such properties has yet to be clarified. Here we demonstrate that monolayer 1$T$-NbTe$_2$ is a new class of two-dimensional TMD which has the star-of-David lattice similarly to bulk TaS$_2$ and isostructural monolayer NbSe$_2$, but exhibits a metallic ground state with an unusual lattice periodicity ($\sqrt{19}{\times}\sqrt{19}$) characterized by the sparsely occupied star-of-David lattice. By using angle-resolved photoemission and scanning-tunneling spectroscopies in combination with first-principles band-structure calculations, we found that the hidden Fermi-surface nesting and associated CDW formation are a primary cause to realize this unique correlated metallic state with no signature of Mott gap. The present result points to a vital role of underlying fermiology to characterize the Mott phase of TMDs. 
\end{abstract}

\maketitle
One of key challenges in materials science is to find outstanding two-dimensional (2D) materials by reducing the dimensionality from bulk (3D) to 2D, as highlighted by the discovery of room temperature quantum Hall effect in graphene \cite{Novoselov}. Transition-metal dichalcogenides (TMDs) offer a fertile platform to explore exotic 2D materials, since two-dimensionalization of bulk TMDs often creates fundamentally different physical properties such as the Ising superconductivity associated with the space-inversion-symmetry breaking in monolayer 1$H$-NbSe$_2$ [ref.2] and the quantum spin Hall insulator phase in monolayer 1$T$'-WTe$_2$ and 1$T$'-WSe$_2$ [refs.3,4]. When electron correlation is introduced into 2D systems, even more exotic quantum states would emerge, as exemplified by the discovery of superconductivity on the verge of Mott-insulating phase in twisted bilayer (BL) graphene, where the enhanced electron correlation due to the band narrowing by moir\'{e} potential plays a key role \cite{Cao}. A 2D Mott-insulating state that coexists with charge-density wave (CDW) in monolayer 1$T$-NbSe$_2$ and TaSe$_2$ has been also discussed to be triggered by the enhanced electron correlation due to the CDW-induced band narrowing \cite{Nakata1,Nakata2,Chen2}.

A central player for such unique CDW-Mott phase is the star-of David (soD) cluster of transition-metal atoms, where the corner atom is slightly displaced from the original position towards the central atom [inset to Fig. 3(d)]. In this soD cluster, the half-filling condition to realize the Mott-insulating phase is satisfied, because, as exemplified in the case of bulk 1$T$-TaS$_2$, twelve electrons at the displaced twelve Ta atoms form 6 fully occupied bands and remaining one electron at the central Ta atom forms a half-filled metallic band \cite{Wilson1,Fazekas1,Thomson,Fazekas2}. However, the mechanism of CDW-Mott phase in TMDs is still far from being well understood \cite{Nakata1,Nakata2,Chen2,Wilson1,Fazekas1,Thomson,Fazekas2,Ang,Ma,Cho,Lee,Wang,Butler,Liu}, owing to the existence of complex energy bands in the CDW phase. It is still unknown how electron correlation, CDW, and Mott phases are interrelated. It is thus highly important to pin down a key ingredient to realize the Mott phase by exploring ultrathin 2D TMDs. In this regard, monolayer 1$T$-NbTe$_2$ is a promising target because it is isostructural to Mott-insulating monolayer 1$T$-NbSe$_2$. Bulk NbTe$_2$ crystallizes in the 1$T$ structure at high temperatures. Below 530 K, it undergoes a structural transition to the monoclinic 1$T$'' phase with the 3$\times$1$\times$3 periodic lattice distortion \cite{Selte,Wilson2,Brown} associated with the CDW triggered by the Fermi-surface (FS) nesting \cite{Battaglia}. Superconductivity was observed below $T_{\rm C}$ = 0.5 K [ref.24]. Despite such intensive studies of bulk properties, basic properties such as crystal structure and ground-state characteristics of monolayer NbTe$_2$ remain unexplored. It is also important to clarify the electronic states of isostructural Se- and Te-based TMDs to understand how the difference in the band character between Se and Te manifests itself in the exotic physical properties such as Mott insulating properties and CDW.

\begin{figure*}
\begin{center}
\includegraphics[width=6.5in]{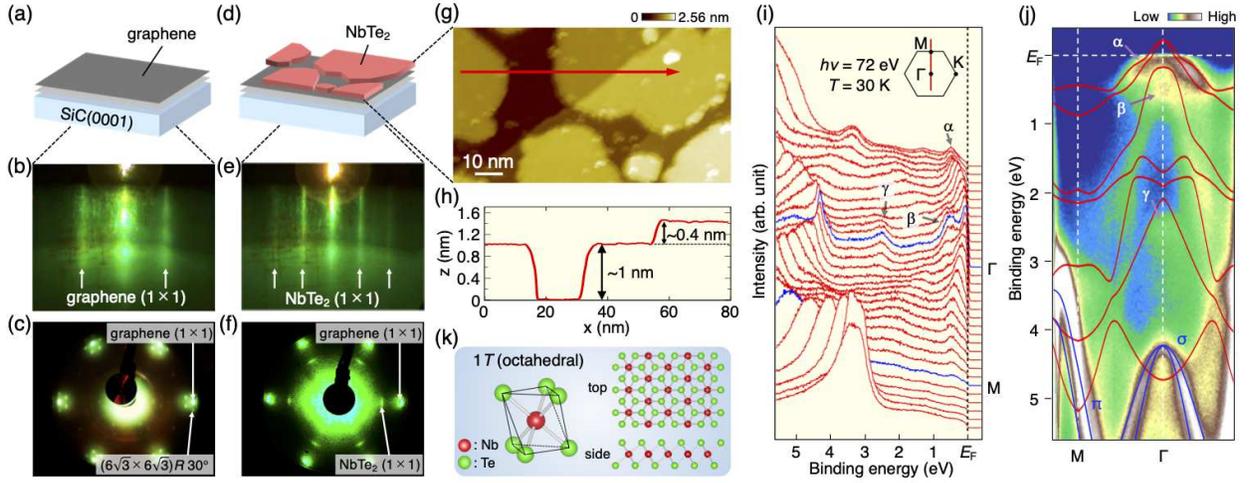}
\caption{(a) Schematic view of BL graphene on SiC. (b),(c) RHEED and LEED patterns of bilayer (BL) graphene. (d)-(f) Same as (a) to (c) but for NbTe$_2$ on BL graphene. (g) Constant-current STM image at $T$ = 4.8 K in the surface area of 56$\times$100 nm$^2$ (sample bias voltage $V_{\rm s}$ = -2 V, and set-point tunneling current $I_{\rm t}$ = 50 pA). (h) Height profile along a cut indicated by red allow in (g). (i) EDCs for monolayer NbTe$_2$ measured at $T$ = 30 K with $h\nu$ = 72 eV along the ${\Gamma}M$ line of NbTe$_2$ Brillouin zone (inset). (j) Plot of valence-band ARPES intensity for monolayer NbTe$_2$, compared with the calculated band structure for monolayer 1$T$-NbTe$_2$ (red curves) and BL graphene (blue curves). Experimental three holelike bands are labeled as $\alpha, \beta$, and $\gamma$ in (i) and (j). (k) Schematic crystal structure of monolayer 1$T$-NbTe$_2$.}
 \end{center}
\end{figure*}

  In this paper, we report angle-resolved photoemission spectroscopy (ARPES) and scanning tunneling microscopy (STM) studies on the electronic state of monolayer 1$T$-NbTe$_2$  fabricated on BL graphene. We uncovered the metallic $\sqrt{19}{\times}\sqrt{19}$ CDW state coexisting with an unusual soD lattice distortion in 1$T$-NbTe$_2$ in stark contrast to the gapped Mott-insulating nature of monolayer 1$T$-NbSe$_2$. We discuss implications of the present results to address the origin of unique CDW-Mott phase in 2D TMDs.
  
  Monolayer NbTe$_2$ films were grown on BL graphene by the molecular-beam-epitaxy (MBE) method. ARPES measurements were carried out using a MBS-A1 electron-energy analyzer (MB Scientific AB) at Tohoku University and a DA-30 electron-energy analyzer (Omicron-Scienta) at beamline BL28 in Photon Factory, KEK. STM measurements were carried out using a custom-made ultrahigh vacuum STM system \cite{Iwaya}. First-principles band-structure calculations were carried out by using the Quantum Espresso code package \cite{Giannozzi} For details, see Appendix \ref{App. A}.

First, we present fabrication and characterization of monolayer NbTe$_2$. To fabricate a NbTe$_2$ film, we used the van der Waals epitaxy technique by using BL graphene grown on silicon carbide as a substrate [Fig. 1(a)]. Figure 1(b) shows the RHEED (reflection high energy electron diffraction) pattern of BL graphene on 6H-SiC(0001) which signifies a 1$\times$1 streak pattern together with a weaker 6$\sqrt{3}{\times}$6$\sqrt{3}R$30$^\circ$ pattern originating from BL graphene and carbon-mesh layer beneath it, respectively. Corresponding spots are also visible in the LEED (low-energy electron diffraction) pattern in Fig. 1(c). After co-evaporation of Nb and Te atoms onto the substrate kept at 300 $^\circ$C under ultrahigh vacuum, the intensity of 6$\sqrt{3}{\times}$6$\sqrt{3}$ spot is reduced and a new 1$\times$1 pattern appears [Figs. 1(d)-(f)]. This behavior is characteristic of TMD ultrathin films, as observed in various monolayer TMDs such as NbSe$_2$ and TaSe$_2$ [refs.6-8]. STM measurements revealed the formation of monolayer NbTe$_2$ islands [Fig. 1(g)] whose height is $\sim$ 1.0 nm [Fig. 1(h)], in rough agreement with the distance between adjacent NbTe$_2$ layers in bulk 1$T$-NbTe$_2$ ($\sim$0.7 nm) \cite{Wilson3}, supporting its monolayer nature (note that another 0.4-nm-height step originates from the step of the SiC substrate).

We characterized the overall band structure of monolayer NbTe$_2$ film by $in$-$situ$ angle-resolved photoemission spectroscopy (ARPES). Figure 1(i) displays the energy distribution curves (EDCs) at $T$ = 30 K measured at $h\nu$ = 72 eV along the ${\Gamma}M$ cut of hexagonal NbTe$_2$ Brillouin zone. Besides the band structure originating from BL graphene situated at the binding energy ($E_{\rm B}$) higher than $\sim$3 eV, one can clearly identify several dispersive features originating from NbTe$_2$ within 3 eV of the Fermi level ($E_{\rm F}$). There exist three holelike bands centered at the $\Gamma$ point; one has a shallow dispersion (labeled as $\alpha$) within 1 eV of $E_{\rm F}$ and appears to cross $E_{\rm F}$ in the vicinity of the $\Gamma$ point, and the other two have a wider dispersion each topped at $E_{\rm B}$ = 0.5 and 2.5 eV at the $\Gamma$ point (labeled as $\beta$ and $\gamma$, respectively). To see more clearly the dispersive features, we plot in Fig. 1(j) the ARPES intensity as a function of $k_{x}$ and $E_{\rm B}$, together with the calculated band structure for monolayer 1$T$-NbTe$_2$ (red) and BL graphene (blue). One can immediately recognize that the intense features at higher $E_{\rm B}$'s well overlap with the calculated $\pi$ and $\sigma$ bands of BL graphene. A qualitative matching can be also found for the NbTe$_2$ bands; the calculated holelike Te 5$p$ bands topped at 0.2 eV and 2.1 eV seem to have experimental counterparts although there exist quantitative differences in their energy positions. The calculated holelike bands crossing $E_{\rm F}$ which are attributed to the Nb 4$d$ (outer) and Te 5$p$ (inner) orbitals also have a good correspondence with the experimental band which crosses $E_{\rm F}$ around the $\Gamma$ point. Such an overall agreement between the experiment and calculation suggests that monolayer NbTe$_2$ on BL graphene takes the 1$T$ structure [Fig. 1(k)], as corroborated by our STM observation of soD clusters (Fig. 3) which are known to be stabilized only in the 1$T$ structure. It is also inferred from a reasonable matching in the energy position of bands around the $\Gamma$ point between experiments and calculations shown in Fig. 1(j) that our film is almost stoichiometric (note that a slight disagreement at around the $\Gamma$ point in higher $E_{\rm B}$'s may be explained in terms of the band renormalization due to the electron correlation). The 1$T$ nature of our NbTe$_2$ film is also supported by the fact that the experimental band structure shows a better agreement with the calculated band structure for the 1$T$ phase than that for the 1$H$ phase (for details, see Appendix \ref{App. B}). Also, the $E_{\rm F}$-crossing of bands in Figs. 1(i) and 1(j) is distinct from the large Mott gap exceeding 0.2 eV in isostructural monolayer 1$T$-NbSe$_2$ and 1$T$-TaSe$_2$ \cite{Nakata1,Nakata2,Chen2}. We will come back to this point later.

\begin{figure}
\begin{center}
\includegraphics[width=3.5in]{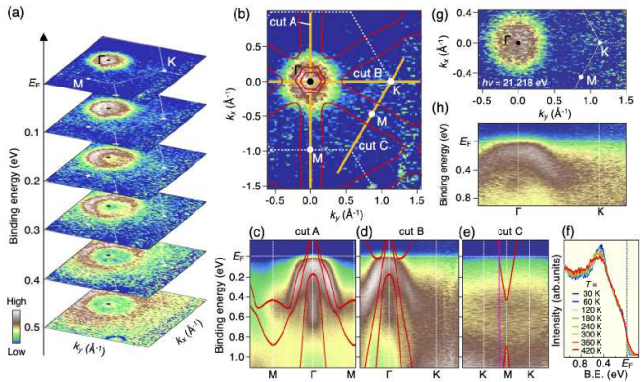}
\caption{(a) ARPES-intensity plots as a function of 2D wave vectors ($k_{x}$ and $k_{y}$) for representative $E_{\rm B}$ slices measured at $T$ = 30 K with $h\nu$ = 72 eV. (b) Same as (a) but at $E_{\rm B}$ = $E_{\rm F}$, overlaid with the calculated Fermi surface (red curves). (c)-(e) ARPES intensity plots near $E_{\rm F}$ for monolayer NbTe$_2$ measured along the ${\Gamma}M$ (cut A), ${\Gamma}K$ (cut B), and $KMK$ (cut C) cuts shown in (b), respectively. Red curves correspond to the calculated band structure for monolayer 1$T$-NbTe$_2$. (f) Temperature dependence of EDC measured with $h\nu$ = 21.218 eV at the $k_{\rm F}$ point of the calculated triangular pocket along the MK cut shown by a puple line in (e). (g), (h) Fermi-surface mapping and ARPES-intensity plot as a function of $k_{y}$ and $E_{\rm B}$, respectively, measured with $h\nu$ = 21.218 eV at $T$ = 440 K for monolayer 1$T$-NbTe$_2$. }
 \end{center}
\end{figure}

Although the experimental band structure around the $\Gamma$ point shows a reasonable agreement with the calculation for monolayer 1$T$-NbTe$_2$, we found a fatal disagreement of the Fermi-surface topology around the $K$ point. The experimental Fermi surface of monolayer 1$T$-NbTe$_2$ is characterized by the existence of a hole pocket only at the $\Gamma$ point. The holelike nature is directly visualized by the equi-energy contour plot at $T$ = 30 K in Fig. 2(a) showing a systematic expansion of the intensity pattern on increasing $E_{\rm B}$. As shown in Fig. 2(b), while the calculated Fermi surface consists of a large triangular hole pocket centered at the $K$ point besides the hexagonal one at the $\Gamma$ point, the spectral weight corresponding to this triangular Fermi surface is missing in the ARPES intensity. To further examine this unusual behavior, we compare the ARPES intensity near $E_{\rm F}$ along several high-symmetry $\textbf{k}$ cuts (cuts A-C) with the corresponding calculated band structure in Figs. 2(c)-2(e). Although the experimental band structure along the ${\Gamma}M$ cut (cut A) shows a reasonable agreement with the calculation, the holelike Te 5$p$ band along the ${\Gamma}K$ cut (cut B) shows a much smaller group velocity and no shallow electron bands exist aside from the hole band. Moreover, along the $MK$ cut (cut C), the spectral feature is broad and flat, and one cannot identify a predicted highly dispersive V-shaped band in the experiment. We have confirmed that the flat feature is robust against the variation in photon energy and light polarization. This suggests that it is not an artifact associated with the matrix-element effect of photoelectron intensity, but is an intrinsic nature of monolayer NbTe$_2$ (for details, see Appendix \ref{App. C}). Temperature-dependent ARPES measurements at the $k_{\rm F}$ point along the MK cut shown in Fig. 2(f) signify a broad peak at $E_{\rm B} \sim$ 0.5 eV which gradually smears out on increasing temperature but still survives at $T$ = 420 K. This peak is likely associated with the formation of CDW (for detailed discussion of its origin, see Appendix \ref{App. D}). We have confirmed that the $K$-centered pocket is absent in a wide temperature range, as highlighted by the Fermi-surface mapping [Fig. 2(g)] as well as the ARPES intensity along the ${\Gamma}K$ cut [Fig. 2(h)] measured at $T$ = 440 K.

\begin{figure*}
\begin{center}
\includegraphics[width=6.4in]{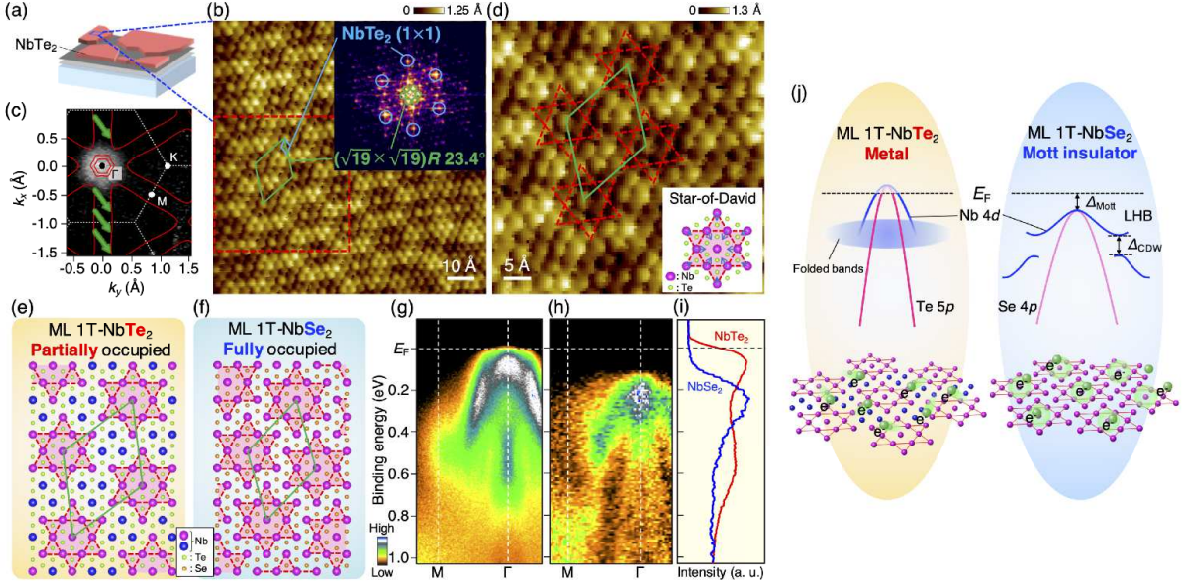}
\caption{(a) Schematics of monolayer NbTe$_2$ islands on BL graphene. (b) STM image in the surface area of 10$\times$10 nm$^2$ ($V_{\rm s}$ = -1.5 V and $I_{\rm t}$ = 0.8 nA). Inset shows Fourier-transform image of (b). (c) Calculated Fermi surface for monolayer 1$T$-NbTe$_2$ (red curves), overlaid on the ARPES-intensity plot. The $\sqrt{19}{\times}\sqrt{19}$ $R$ 23.4$^\circ$ nesting vector is shown by green arrows. (d) STM image magnified in the area shown by red square in (b). $V_{\rm s}$ and $I_{\rm t}$ were set to be -1.0 V and 0.8 nA, respectively. Inset shows the illustration of soD cluster. (e),(f) Comparison of the schematic soD lattice between monolayer NbTe$_2$ and NbSe$_2$. (g),(h) Near-$E_{\rm F}$ ARPES intensity along the ${\Gamma}M$ cut for monolayer NbTe$_2$ and NbSe$_2$, respectively, measured at $T$ = 30 K. (i) EDC at the $\Gamma$ point for monolayer NbTe$_2$ and NbSe$_2.$ (j) Schematics on the relationship between the band dispersion and the soD lattice for monolayer NbTe$_2$ and NbSe$_2$.}
 \end{center}
\end{figure*}

To clarify the mechanism behind the disappearance of a large triangular pocket, we have performed high-resolution STM measurements on a NbTe$_2$ island [Fig. 3(a)]. The STM image at $T$ = 4.8 K in Fig. 3(b) signifies individual Te atoms originating from the top layer of monolayer NbTe$_2$. Noticeably, the atomic image exhibits a strong intensity modulation; a dark region involving at least 7 Te atoms arranges periodically and forms a hexagonal superlattice surrounded by a brighter region with a honeycomb-shaped pattern. The Fourier-transformation image in the inset to Fig. 3(b) signifies $\sqrt{19}{\times}\sqrt{19}R$23.4$^\circ$ superspots (green circles) which correspond to the supercell shown by green rhombus in Fig. 3(b). We found in Fig. 3(c) that the $\sqrt{19}{\times}\sqrt{19}R$23.4$^\circ$ superlattice vector well connects the parallel segments of calculated triangular pockets at the K point, consistent with the calculated electronic susceptibility (see Appendix \ref{App. E}) [note that the crystal geometry allows the existence of two types of crystal domains ($R{\pm}$23.4$^\circ$) which equally satisfy the nesting condition, as detailed in Appendix \ref{App. F}]. These suggest the formation of CDW triggered by the Fermi-surface nesting, which is also inferred from the observation of real-space charge reversal across the CDW gap similarly to the case of KV$_3$Sb$_5$ in the CDW phase \cite{Jiang} (see Appendix \ref{App. G}). The CDW may fully gap out the triangular pocket and create many backfolded subbands. Electron correlation may further smear out the fine structure of each subband, leading to the featureless spectral intensity around the K point as seen in Fig. 2(e). The CDW origin of disappearance of the triangular pocket needs to be experimentally confirmed by modulating the FS-nesting condition by carrier doping, electrical gating, or epitaxial strain. We have estimated the CDW transition temperature ($T_{\rm CDW}$) to be much above 420 K from the persistence of a peak in the EDC in Fig. 2(f) (note that it was difficult to estimate $T_{\rm CDW}$ from the LEED pattern because superspots were not clearly seen. Such a vague feature of LEED superspots despite the CDW formation was also recognized in some other monolayer TMDs \cite{Ryu, Coelho}).

To obtain further insights into the characteristics of CDW, we show in Fig. 3(d) a magnified STM image. One may see that the dark region consists of twelve atoms which coincide well with the schematic double triangles rotated by 60$^\circ$ from each other (red broken triangles), a signature of soD cluster (inset) reported in bulk 1$T$-TaS$_2$ [refs.11,14,15] as well as monolayer 1$T$-NbSe$_2$ and 1$T$-TaSe$_2$ [refs.6-8]. These soD clusters form a hexagonal lattice with the $\sqrt{19}{\times}\sqrt{19}R$23.4$^\circ$ periodicity. As a consequence, residual atoms outside the soD clusters are seen as relatively bright spots intervening adjacent soD clusters. It is thus inferred that although isostructural family of monolayer NbTe$_2$, NbSe$_2$, and TaSe$_2$ (together with bulk 1$T$-TaS$_2$) commonly form the soD cluster, that for NbTe$_2$ partially occupies the lattice [Fig. 3(e); 13 atoms among 19 atoms in the superstructure unit cell are involved in the soD cluster], in stark contrast to the full occupation in other TMDs [Fig. 3(f)] \cite{Nakata1,Nakata2,Chen2,Thomson,Ma,Cho}, demonstrating a unique characteristic of monolayer NbTe$_2$. It is worthwhile to note that, although the $\sqrt{19}{\times}\sqrt{19}R$23.4$^\circ$ periodicity is absent in bulk NbTe$_2$, it was reported to locally emerge when the bulk sample was pulse heated by electron beam \cite{Landuyt}, and such periodicity was discussed in terms of the CDW driven by the FS nesting \cite{Wilson3}. In this respect, it would be reasonable to infer that the soD modulation observed in monolayer is associated with the formation of CDW, whereas this point needs to be clarified by further experiments.

Now we discuss the relationship between the soD lattice and Mott characteristics. We suggest that the occupation of soD clusters is crucial for understanding the electronic properties at low temperature. Monolayer NbSe$_2$ is an insulator as seen from the spectral-weight suppression around $E_{\rm F}$ in the ARPES intensity [Fig. 3(h)] and the apparent energy-gap opening in the EDC at the $\Gamma$ point [Fig. 3(i)]. This insulating gap was attributed to the Mott-Hubbard gap \cite{Nakata1}, because the half-filling condition is satisfied for each soD cluster and there are no other conducting electrons because of the fully occupied nature of the soD lattice. On the other hand, in NbTe$_2$, half-filling condition for the Nd 4$d$ orbital is globally violated because the Te 5$p$ band participates in the Fermi surface. Also, electron hopping between the soD and outside-soD regions could disturb the effective half-filling nature in a single soD cluster and deteriorate the electron localization. These lead to the metallic nature as seen from the sizable spectral weight at $E_{\rm F}$ in the EDC at the $\Gamma$ point [see Fig. 3(i)].

The observed intriguing spectral difference between monolayer NbTe$_2$ and NbSe$_2$ is also explained in terms of fermiology. As shown in Fig. 3(c), the calculated Fermi surface in the normal state for monolayer NbTe$_2$ consists of two hole pockets at the $\Gamma$ point associated with the $E_{\rm F}$-crossing of the topmost Nb 4$d$ and Te 5$p$ bands [schematically shown in Fig. 3(j)], besides a large triangular pocket centered at the $K$ point due to the Nb 4$d$ band. On the other hand, these small pockets at the $\Gamma$ point are absent in the calculation for monolayer NbSe$_2$ [ref.32] because the Se bands are pulled downward and fully occupied [Fig. 3(j)], as in the case of other chalcogenides \cite{Nakata2,Umemoto}. Since the total volume of Fermi surface must be identical between NbTe$_2$ and NbSe$_2$ according to the Luttinger theorem, such a difference causes the change in the volume of Nd 4$d$ pocket and the resultant modification in the Fermi-surface topology \cite{Kikuchi}. This argument is also supported by estimating the carrier number in the observed Fermi surface. The result indicates 1.1 electrons/unit-cell for the Nd-4$d$ triangular pocket in NbTe$_2$ while it is 1.0 electrons/unit-cell for NbSe$_2$. The deviation from the half-filling condition in NbTe$_2$ is ascribed to the extra Te 5$p$ hole carriers. Such a difference in the fermiology leads to the difference in the Fermi-surface-nesting vector, i.e. $\sqrt{19}{\times}\sqrt{19}R$23.4$^\circ$ for NbTe$_2$ and $\sqrt{13}{\times}\sqrt{13}R$13.9$^\circ$ for NbSe$_2$. Since the small pockets at the $\Gamma$ point in NbTe$_2$ are not well connected with each other by the nesting vector, a gapless $\textbf{k}$ region remains in the CDW phase, contributing to the absence of Mott characteristics of NbTe$_2$ in contrast to NbSe$_2$ where the small pocket is intrinsically absent. In this regard, it is very important to systematically fabricate monolayer NbSe$_{2-x}$Te$_x$ films and investigate the Mott-transition characteristics as a function of $x$.

One may wonder how the above interpretation based on the real space (occupation of soD) is reconciled with that of the momentum-space (fermiology). The small hole pocket at the $\Gamma$ point seen by ARPES may be attributed to the outside-soD region observed by STM.  However, this argument is too simplistic because the electron localization in the real space is incompatible with the formation of energy band that requires non-local nature in the real space. This in return suggests that the soD and outside-soD regions are not phase separated and electrons can coherently hop between these two regions. This scenario seems consistent with the $dI/dV$ curves in the STM data which smoothly evolve across the boundary of two regions (see Appendix \ref{App. H}). In this context, it is inferred that a small but finite atomic displacement takes place even in the outside-soD region. A sophisticated diffraction study is necessary to clarify this point.

Now we discuss the difference between the metallic NbTe$_2$ and Mott-insulating NbSe$_2$ in terms of the effective electron correlation. Since the on-site Coulomb energy $U$ for Nb 4$d$ electrons is expected to be similar between NbSe$_2$ and NbTe$_2$, a crucial parameter to determine the effective coulomb energy $U/W$ is the bandwidth $W$ in the CDW phase. Since the reconstructed Brillouin zone for NbTe$_2$ is smaller than that for NbSe$_2$ due to the larger superlattice unit cell ($\sqrt{19}{\times}\sqrt{19}$ vs $\sqrt{13}{\times}\sqrt{13}$), the reconstructed subbands are expected to be flatter in NbTe$_2$ than in NbSe$_2$ when one assumes that the strength of CDW potential is similar. This leads to smaller $W$ and resultantly larger $U/W$ in NbTe$_2$ suggestive of the stronger ``Mottness'' in NbTe$_2$. However, this is opposite to our observation, suggesting a less important role of electron correlation to account for the difference between NbTe$_2$ and NbSe$_2$. It is noted though that this conjecture is not fully supported by the current experiment because each subband is hard to resolve due to the strong spectral broadening. Another alternative explanation to account for the difference between the metallic NbTe$_2$ and insulating NbSe$_2$ is that the Te 5$p$ band with wide bandwidth (which is even wider than that of bulk 1$T$-TaS$_2$ in the nearly commensurate CDW phase \cite{Ang}) crosses $E_{\rm F}$ and reduces the effective electron correlation to prevent this system from the Mott transition.

In conclusion, we have performed ARPES and STM combined with first-principles band calculations to study the electronic structure of monolayer 1$T$-NbTe$_2$ which is characterized by a partially occupied soD lattice possibly associated with the CDW formation. The present observation demonstrates that the formation of soD lattice is not a sufficient condition to realize the Mott-insulating phase which coexists with CDW, while it has been discussed as a necessary condition in bulk 1$T$-TaS$_2$ and monolayer TMDs. We also found that the underlying Fermi-surface topology and associated hidden Fermi-surface nesting play a crucial role to control the key electronic properties such as periodicity of superlattice, occupation of soD lattice, and Mott-insulating vs metallic properties. The present result opens a pathway toward switching and controlling the Mott-insulating and CDW phases via fermiology engineering in ultrathin TMDs.

\begin{acknowledgments}
We thank Y. Nakata, M. Kitamura, K. Horiba, and H. Kumigashira for their help in the ARPES experiments. This work was supported by JST-CREST (no. JPMJCR18T1), JST-PRESTO (no. JPMJPR20A8), Grant-in-Aid for Scientific Research on Innovative Areas ``Topological Materials Science'' (JSPS KAKENHI Grant numbers JP15H05853, and JP15K21717), Grant-in-Aid for Scientific Research on Innovative Areas ``Discrete Geometric Analysis for Materials Design'' (JSPS KAKENHI Grant numbers 20H04624), Grant-in-Aid for Scientific Research (JSPS KAKENHI Grant numbers JP18H01821, and JP21H01757), Science research projects from Shimadzu Science Foundation, and World Premier International Research Center, Advanced Institute for Materials Research. T. Kawakami and T. Kato acknowledges support from GP-Spin at Tohoku University.
\end{acknowledgments}

\newpage
\bibliographystyle{prsty}

\appendix
\section{Sample fabrication, ARPES measurements, and band calculations}\label{App. A}

High-quality monolayer NbTe$_2$ films were grown on bilayer (BL) graphene by the molecular-beam epitaxy (MBE) method in an ultrahigh vacuum of 5$\times$10$^{-10}$ Torr. BL graphene was grown by annealing an $n$-type Si-rich 6H-SiC(0001) single-crystal wafer \cite{Nakata1,Nakata2} by resistive heating at 1100 $^\circ$C for 30 min in an ultrahigh vacuum better than 1.0$\times$10$^{-9}$ Torr. Monolayer NbTe$_2$ film was grown by co-evaporating Nb and Te on the BL graphene substrate. The substrate temperature was kept at 300 $^\circ$C during the epitaxy. The as-grown film was annealed at 300 $^\circ$C for 30 min to improve the crystallinity, and then transferred to the ARPES-measurement chamber without breaking vacuum. The growth process was monitored in real-time by reflection high-energy electron diffraction (RHEED). 

Scanning tunneling microscopy (STM) measurements were carried out using a custom-made ultrahigh vacuum STM system \cite{Iwaya}. Te capping ($\sim$ 10 nm thick) for surface protection of NbTe$_2$ film was removed in the STM chamber by Ar$^+$ ion sputtering for 30 min and annealing at 250 $^\circ$C for 30 min. STM measurements were carried out with an electrochemically etched W tip at 4.8 K under UHV below 2.0$\times$10$^{-10}$ Torr. All STM images were obtained in constant current mode.
ARPES measurements were carried out using a MBS-A1 electron-energy analyzer with a high-flux helium discharge lamp at Tohoku University and a DA-30 electron energy analyzer with synchrotron radiation at the beamline BL-28A at Photon Factory (KEK). The He I$\alpha$ resonance line ($hv$ = 21.218 eV) and linearly polarized light of $hv$ = 72 eV were used to excite photoelectrons. The energy and angular resolutions were set to be 16-30 meV and 0.2-0.3 $^\circ$C, respectively. The Fermi level ($E_{\rm F}$) of samples was referenced to that of a gold film deposited onto the sample substrate. 

First-principles band-structure calculations for free-standing monolayer 1$T$- and 1$H$-NbTe$_2$ were carried out by using the Quantum-Espresso code \cite{Giannozzi}. Spin-orbit interactions were included in the calculation. The plane-wave cutoff energy and the k-point mesh were set to be 70 Ry and 12$\times$12$\times$1, respectively. Thickness of the inserted vacuum layer for monolayer was set to be $\sim$ 10 $\rm{\AA}$.

\section{Comparison between ARPES intensity and band calculations for monolayer 1$T$/1$H$-NbTe$_2$}\label{App. B}

\begin{figure}
\begin{center}
\includegraphics[width=3.5in]{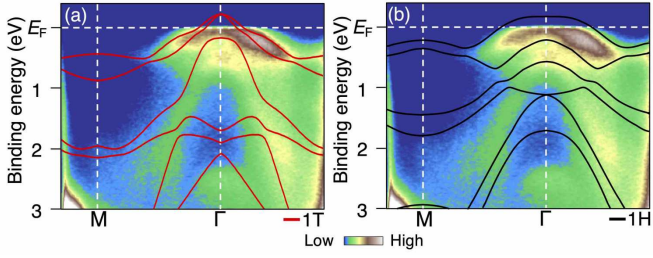}
\caption{(a), (b) ARPES intensites of monolayer NbTe$_2$ compared with the calculated band dispersion for monolayer 1$T$-NbTe$_2$ (red curves) and monolayer 1$H$-NbTe$_2$ (black curves), respectively.}
 \end{center}
\end{figure}

To clarify the possibility of different crystal structures, we have carried out band-structure calculations for monolayer NbTe$_2$ for 1$T$ and 1$H$ phases. Figures 4(a) and 4(b) show the APRES intensity along the $\Gamma$M cut overlaid by the calculated band structure for monolayer 1$T$- and 1$H$-NbTe$_2$, respectively [note that the calculation does not take into account the formation of star-of-David (soD) lattice]. One can see an overall agreement between experiment and calculation for the 1$T$ case, such as a weakly dispersive holelike band at $E_{\rm B}$ = $E_{\rm F}$-0.1 eV and a highly dispersive holelike band at $E_{\rm B}$ = 0.3-1.5 eV, whereas the matching between experiment and calculation is relatively poor for the 1$H$ case [Fig. 4(b)]. This supports the 1$T$ nature of our NbTe$_2$ film.

\section{Photon-energy and light-polarization dependences of band structure in monolayer NbTe$_2$}\label{App. C}

\begin{figure}
\begin{center}
\includegraphics[width=3.0in]{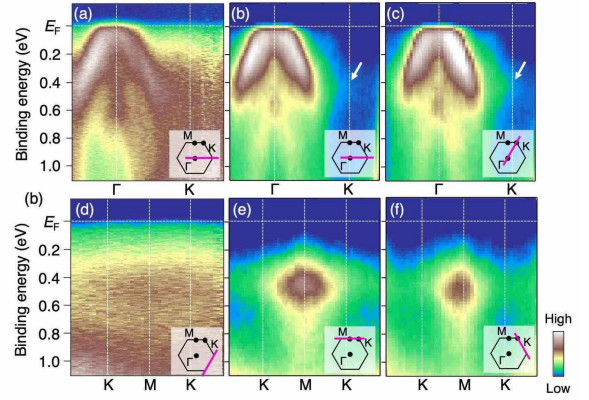}
\caption{(a), (b) ARPES intensities along the $\Gamma$$K$ cut obtained at $hv$ = 21.218 eV with linear horizontal light and at $hv$ = 75 eV with right circular polarized light, respectively. (c) Same as (b) but obtained along another $\Gamma$$K$ cut rotated by 60$^\circ$  with respect to the $\Gamma$$K$ cut in (b). (d)-(f), Same as (a)-(c) cut, but measured along the $MK$ cut. White arrows in (b) and (c) indicate the location of flat feature.}
 \end{center}
\end{figure}

We performed ARPES measurements along the $\Gamma$$K$ and $KMK$ cuts with different sample-light geometry and photon energy ($hv$) to clarify the influence from the matrix-element effect of photoelectron intensity to the ARPES intensity. Figures 5(a) and 5(b) show the ARPES intensity along the $\Gamma$$K$ cut at $hv$ = 21.218 eV with linear horizontal polarization and at $hv$ = 75 eV with right circular polarization, respectively. At $hv$ = 21.218 eV [Fig. 5(a)], one can recognize a broad non-dispersive feature around the $K$ point at $E_{\rm B}$ of $\sim$ 0.5 eV, in addition to a highly dispersive holelike band centered at the $\Gamma$ point. This flat feature is likely associated with the formation of CDW. As shown in Fig. 5(b), a similar flat feature is also observed at $hv$ = 75 eV around the $K$ point (indicated by white arrow), whereas its intensity is largely suppressed. As shown in Fig. 5(c), such a weak and flat feature also exists along another $\Gamma$$K$ cut rotated by 60$^\circ$ with respect to the $\Gamma$$K$ cut in Fig. 5(b). We find that the flat feature is also recognized along the $MK$ cut with different $hv$'s and light polarizations, as shown in Figs. 5(d)-5(f). These results suggest that the existence of observed flat feature (i.e. the absence of a triangular pocket predicted by the DFT calculation) is robust against the variation in photon energy and light polarization. This suggests that the flat feature is not an experimental artifact associated with the matrix-element effect of photoelectron intensity, but is an intrinsic feature of monolayer 1$T$-NbTe$_2$.

\section{Origin of a flat band around the $K$ point for monolayer NbTe$_2$}\label{App. D}

We discuss the origin of a flat feature at $E_{\rm B}$ $\sim$ 0.5 eV around the $K$ point observed in Fig. 2 and Fig. 5. We have considered some possible mechanisms unrelated to the CDW, such as (i) the band gap, (ii) the Anderson gap, and (iii) the Mott-Hubbard gap. The possibility (i) is ruled out because the DFT calculation of monolayer NbTe$_2$ fails to reproduce the flat band and a finite energy gap. The possibility (ii) is also unlikely because our NbTe$_2$ film is close to stoichiometry and the ARPES spectrum around the K point (Fig. 2 of the main text) does not show a power-low behavior against $E_{\rm B}$ expected from the disorder-induced Coulomb gap. The possibility (iii) is also ruled out because the gap size $\Delta$ of $\sim$ 0.5 eV is too large compared to the Mott-Hubbard gap of isostructural 1$T$-NbSe$_2$ ($\Delta$ $\sim$ 0.1 eV) despite the expectation of a similar on-site Coulomb energy. Taking into account these points, we think that the flat band is associated with the CDW gap formation.

\section{Calculated electronic susceptibility for monolayer NbTe$_2$}\label{App. E}

\begin{figure}
\begin{center}
\includegraphics[width=1.5in]{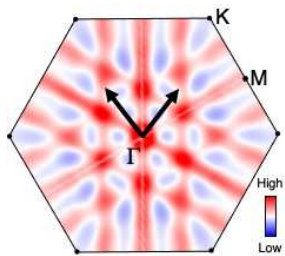}
\caption{Intensity plot of the calculated imaginary part of electronic susceptibility Im$\chi$(\textbf{q}) for monolayer 1$T$-NbTe$_2$. Arrows correspond to \textbf{q} = 1/$\sqrt{19}$ \textbf{G} nesting vectors.}
 \end{center}
\end{figure}

To clarify the origin of CDW in monolayer NbTe$_2$, we calculated the electronic susceptibility $\chi$(\textbf{q}) based on the calculated Fermi surface shown in Fig. 2(b). Figure 6 shows the imaginary part of calculated electronic susceptibility [Im$\chi$(\textbf{q})] for monolayer 1$T$-NbTe$_2$. One can recognize three straight lines running along the $\Gamma$M direction. Same lines also appear along the other two equivalent $\Gamma$M directions rotated by $\pm$ 60$^\circ$  from each other. These lines are mainly associated with the parallel segments of the calculated triangular pockets. One can see that two types of nesting vectors \textbf{q} for $\sqrt{19}{\times}\sqrt{19}R+$23.4$^\circ$ and $\sqrt{19}{\times}\sqrt{19}R-$23.4$^\circ$ superstructures suggested from the STM data are located near the intersection of straight lines at which the Im$\chi$(\textbf{q}) value shows an overall enhancement. This supports the Fermi-surface-instability-driven CDW in monolayer NbTe$_2$. It is noted that the electronic susceptibility does not show sharp peak or strong divergence at the corresponding nesting vector. Similar behavior was also recognized in other 2D TMDs, such as 1$T$-VTe$_2$ and 2$H$-NbSe$_2$ \cite{Kawakami,Soumyanarayanan,Johannes}.

\section{Two types of crystal domains in monolayer NbTe$_2$ on bilayer graphene}\label{App. F}

\begin{figure}
\begin{center}
\includegraphics[width=3.4in]{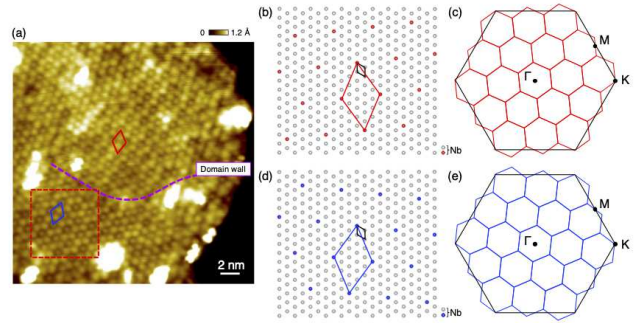}
\caption{(a), STM image at $T$ = 4.8 K in the surface area of 25$\times$25 nm$^2$ on NbTe$_2$ island ($V$$_s$ = -1.6 V, and $I$$_t$ = 0.1 nA). (b) Real-space configuration of Nb atoms (open circles) and superlattice unit cell (red rhombus) for the $\sqrt{19}{\times}\sqrt{19}R-$23.4$^\circ$ superstructure and (c) corresponding reconstructed Brillouin zone (red) drawn on the original 1$\times$1 Brillouin zone (black). (d), (e) Same as (b) and (c), but for the $R+$23.4$^\circ$ superstructure.}
 \end{center}
\end{figure}

We found two types of domain structures associated with the $\sqrt{19}{\times}\sqrt{19}R$23.4$^\circ$ star-of-David (soD) lattice by STM measurements. Figure 7(a) shows the STM image of monolayer 1$T$-NbTe$_2$ island on bilayer graphene. One can identify existence of two types of superstructures. One has a $\sqrt{19}{\times}\sqrt{19}R-$23.4$^\circ$ periodicity (red rhombus) and another is rotated by 13.2$^\circ$ with respect to the former one (blue rhombus). The latter domain is assigned to the superstructure with $\sqrt{19}{\times}\sqrt{19}R+$23.4$^\circ$ by taking into account the 6-fold symmetry of NbTe$_2$ (i.e. 13.2$^\circ$ = 60$^\circ$ - 23.4$^\circ$ - 23.4$^\circ$). The existence of two types of domains is naturally expected from the real-space configuration of Nb atoms and the supercell [Figs. 7(b) and 7(d)] as well as the corresponding Brillouin zones [Figs. 7(c) and 7(e)]. At the boundary between the $R+$23.4$^\circ$ and $R-$23.4$^\circ$ domains, a domain wall is recognized [purple dashed curve of Fig. 7(a)].

\section{STM images across $E_{\rm F}$ for monolayer NbTe$_2$}\label{App. G}

\begin{figure}
\begin{center}
\includegraphics[width=3.4in]{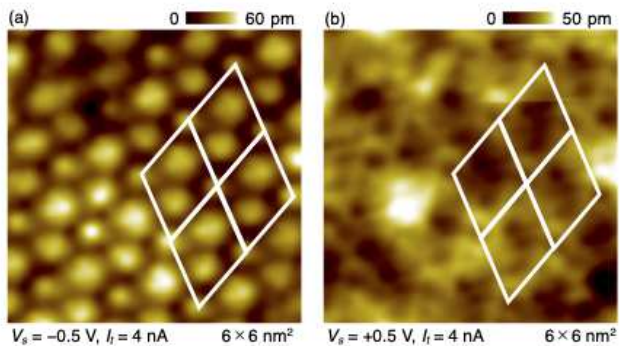}
\caption{(a), (b) STM images at $T$ = 4.8 K in the surface area of 6$\times$6 nm$^2$ measured at (a) negative ($V$$_s$ = -0.5 V, $I$$_t$ = 4 nA) and (b) positive ($V$$_s$ = +0.5 V, $I$$_t$ = 4 nA) bias voltage. White rhombus corresponds to the $\sqrt{19}{\times}\sqrt{19}R$23.4$^\circ$ superstructure unit cell.}
 \end{center}
\end{figure}

We performed STM measurements at negative and positive bias voltages across the CDW gap for monolayer NbTe$_2$. As clearly seen in Fig. 8, the high electron-density region characterized by bright-intensity spots at negative bias voltage [Fig. 8(a)] turns into the low electron-density region at positive bias voltage [Fig. 8(b)], similarly to the case of KV$_3$Sb$_5$ [ref.8]. This supports the real-space charge reversal and the CDW origin of the observed V-shaped gap for monolayer 1$T$-NbTe$_2$.

\section{Local density of states for monolayer NbTe$_2$}\label{App. H}

\begin{figure}

\begin{center}

\includegraphics[width=3.2in]{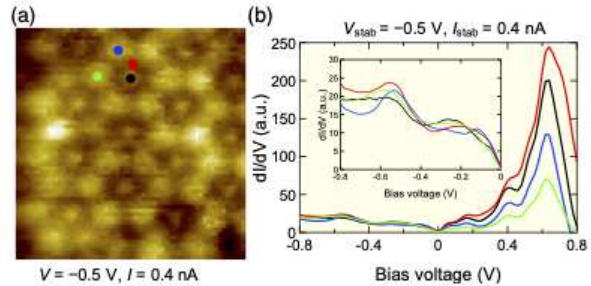}
\caption{(a), STM image at $T$ = 4.8 K in the surface area of 6$\times$6 nm$^2$for monolayer NbTe$_2$. (b) d$I$/d$V$ curves measured at four representative positions inside the soD cluster [red and black filled circles in (a)] and outside it (blue and green ones). Correspondence between the measurement position and d$I$/d$V$ curve is shown by coloring. Inset shows an expansion in the negative bias-voltage region. d$I$/d$V$ spectra were recorded with an open feedback loop using a lock-in technique with a modulation voltage of 20 mV at 987 Hz. The setpoints of voltage and current (Vstab and Istab, respectively) to stabilize the tip before opening feedback loop are indicated in (b).}
 \end{center}
\end{figure}

Figure 9(a) shows the STM image at $T$ = 4.8 K in the surface area of 6$\times$6 nm$^2$ for monolayer NbTe$_2$. Filled circles represent the measurement position of d$I$/d$V$ curves inside the soD cluster (red and black circles) and outside it (blue and green circles). Figure 9(b) shows the d$I$/d$V$ curves for monolayer NbTe$_2$ measured at $T$ = 4.8 K. The color of curves shows the correspondence to the measurement position in Fig. 9(a). One can recognize several peak features in both positive and negative bias voltages irrespective of the location of surface, and there is no abrupt change in the d$I$/d$V$ curve across the boundary between the inside- and outside-soD regions. Expanded image in the negative-bias voltage region in the inset to Fig. 9(b) signifies broad humps at around -0.2 eV and -0.55 eV, which are associated with the Te 5$p$ bands at the $\Gamma$ point and the flat band at $\sim$0.5 eV around the K point observed by ARPES, respectively [Figs. 2(c) and 2(d)]. One can also see the V-shaped density of states (DOS) at $E_{\rm F}$ supportive of the absence of an insulating energy gap and appearance of a soft gap, in stark contrast to a large Mott gap in monolayer 1$T$-NbSe$_2$ and 1$T$-TaSe$_2$ [ref. 6,7,8]. Interestingly, the V-shaped DOS resembles that at the boundary between the metallic and Mott-insulating domains in bulk 1$T$-TaS$_2$ [ref.14,15]. The soft-gap nature with residual DOS at $E_{\rm F}$ seen by STM is also consistent with the ARPES data showing a finite Fermi-edge cut-off associated with the Te 5$p$ band around the $\Gamma$ point seen in Fig. 2.

\end{document}